\documentclass[a4paper,11pt]{article}

\usepackage{amsmath}
\usepackage{amsfonts}
\usepackage{amssymb}
\usepackage{bbm}
\usepackage{pstricks}
\usepackage{pst-plot}
\usepackage{pst-node}
\usepackage{multido}
\usepackage[all]{xy}
\usepackage{rotating}
\usepackage[dvips]{epsfig}
\usepackage{colortbl}
\usepackage{wrapfig}
\usepackage[hang,nooneline]{subfigure}
\usepackage[footnotesize,ruled]{caption}

\usepackage[utf8]{inputenc}

\usepackage{euscript}

\begin{document}

\title{Questioning Newton's second law: What is the structure of equations of motion?}
\author{Claus L\"ammerzahl and Patricia Rademaker \\
ZARM, University of Bremen, Am Fallturm, 28359 Bremen, Germany}

\maketitle

\begin{abstract}
Interactions are explored through the observation of the dynamics of particles. On the classical level the basic underlying assumption in that scheme is that Newton's second law holds. Relaxing the validity of this axiom by, e.g., allowing for higher order time derivatives in the equations of motion would allow for a more general structure of interactions. We derive the structure of interactions by means of a gauge principle and discuss the physics emerging from equations of motion of higher order. One main result is higher order derivatives induce a {\it zitterbewegung}. As a consequence the main motion resulting from the second order equation of motion is rather robust against modifications. The gauge principle leads to a gauge field with the property of a space metric. We confront this general scheme with experimental data. 
\end{abstract}

\section{Introduction}

The most basic approach to the mathematical description of nature is provided by Newton axioms \cite{Newton1872}. The most important of them state that (i) there are inertial systems, (ii) the acceleration of a body with respect to an inertial system is given by 
\begin{equation}
m \ddot{\mbox{\boldmath$x$}} = \mbox{\boldmath$F$}(\mbox{\boldmath$x$}, \dot{\mbox{\boldmath$x$}}, t) \, , \label{Newton3axiom}
\end{equation}
where $m$ is the (inertial) mass of the body, $\ddot{\mbox{\boldmath$x$}}$ its acceleration and $\mbox{\boldmath$F$}$ the force acting on the body which may depend on the position and the velocity of the particle, and that (iii) {\it actio} equals {\it reactio}. 

Leaving aside the fundamental and still unresolved problem of how to really define an inertial system (see, e.g., \cite{SchroeterSchelb95}), the second Newton axiom is a tool to explore the forces and, thus, to measure the fields acting on particles. The electromagnetic field, for example, can be explored and measured through the observation of the acceleration of charged particles under different conditions (different initial conditions, different charges, etc.). Thus, by means of the dynamics of the form \eqref{Newton3axiom} the electromagnetic interaction and, in principle, all other interactions are defined. In many cases one uses quantum equations of motion like the Schr\"odinger or Dirac equation which, via the path integral approach or the Ehrenfest theorem, for example, are also based on \eqref{Newton3axiom}. All these considerations extend to relativistic equations of motion. 

From this observation it is clear that what one defines as interaction or as the corresponding force field depends on the basic structure of the equation of motion. If, for example, the equation of motion is of higher than second order in the time derivative, then interactions could be introduced in a different way and, thus, can have a different structure as we will show below. As a consequence, it is very important to have an {\it experimental} basis for the fundamental equation of motion to be of second order in the time derivative. 

Equations of motion of higher order are related to a different initial value problem: one needs more initial values beyond the initial position and initial velocity. In physical terms this means that with respect to the time coordinate the equation of motion is more non--local than the ordinary second order equations. The extreme case that (formally) an infinite number of initial values are needed is related to equations of motion with memory as, e.g., generalized Langevin equations where the force equation possesses an additional term of the form $\int_0^t C(t - t^\prime) \dot x(t^\prime) dt^\prime$, see, e.g., \cite{Kubo66}. 

Also within quantum gravity scenarios it might be expected that the effective equation of point particles may contain small higher order time derivatives. In fact, since space--time fluctuations in the sense of, e.g., fluctuations of the metrical tensor, yield Langevin--like equations of motion also quantum gravity scenarios naturally are expected to lead to effectively higher order equations of motion where the additional higher derivatives probably scales with, e.g., the Planck length. 

Higher order derivatives in equations of motion occur in effective equations which take back reaction into account. One example for that is the Abraham--Lorentz equation for charged particles taking into account the electromagnetic waves radiated away \cite{Jackson98} or the radiation damping equation in gravity where the emitted gravitational waves are taken into account, \cite{Chiconeetal01}. In the electromagnetic case the leading term is a third time derivative which leads to unphysical runaway solutions what still is an unresolved problem. However, these equations discussed in relation with radiation damping are no fundamental equations, they are effective equations emerging from the fact that one no longer regards the particles as test particles. Here we are only interested in the fundamental equations of motion. 

Since spin is some element of non--locality it is not astonishing that also the dynamics of spinning particles effectively can be described by means of a higher order theory \cite{DamourSchaefer91}. 

Pais and Uhlenbeck \cite{PaisUhlenbeck50} studied higher order mechanical models as a toy model for discussing properties of higher order field theories where higher order derivatives came in naturally in noncommutative models \cite{DouglasNekrasov01} or are introduced in order to eliminate divergences, see, e.g., \cite{Simon90}. Recently, it has been shown in \cite{Dectoretal08} that the energy of the Pais--Uhlenbeck oscillator is bounded from below, is unitary and is free of ghosts, see also \cite{BolonekKosinski07} for further studies in this direction.

In the following we will consider {\it fundamental} higher order equations of motion which can be derived from a variational principle. In order to be able to confront these modified equations of motion with experimental data one first has to investigate the structure of interactions. This will be done by using a gauge principle for a second order Lagrange formalism (which in principle can be applied to Lagrangians of all orders). The solutions of the equations of motion coupled to these generalized gauge fields show that essentially only a {\it zitterbewegung} is introduced by the higher order derivatives showing that the standard equations of motion are rather robust against addition of higher order terms. We also discuss the experimental possibilities to search for effects related to higher order time derivatives. As an interesting by--product of the corresponding higher order gauge formalism we obtain the standard spatial metric as a gauge field. 

\section{Lagrange formalism}

First we will use the Lagrange formalism in order to introduce interactions into a theory containing higher order derivatives. We introduce interactions by means of a gauge principle. We set up our notation by repeating shortly the standard first order formalism and then apply a gauge principle to a second order Lagrange formalism. 

\subsection{First order formalism}

A Lagrange function of first order
\begin{equation}
L = L_0(t, \mbox{\boldmath$x$}, \dot{\mbox{\boldmath$x$}})
\end{equation}
yields the Euler--Lagrange equation of motion
\begin{equation}
0 = \mbox{\boldmath$\nabla$} L - \frac{d}{dt} {\mbox{\boldmath$\nabla$}}_{\dot x} L \, .
\end{equation}
We obtain the same equation of motion from two Lagrange functions if they differ by a total time derivative only
\begin{equation}
L(t, \mbox{\boldmath$x$}, \dot{\mbox{\boldmath$x$}}) \rightarrow L^\prime(t, \mbox{\boldmath$x$}, \dot{\mbox{\boldmath$x$}}) = L_0(t, \mbox{\boldmath$x$}, \dot{\mbox{\boldmath$x$}}) + \frac{d}{dt} f(t, \mbox{\boldmath$x$}) \, ,
\end{equation}
where the function $f$ is allowed to depend on $t$ and $\mbox{\boldmath$x$}$ only. We can expand the total time derivative
\begin{equation}
L^\prime(t, \mbox{\boldmath$x$}, \dot{\mbox{\boldmath$x$}}) = L_0(t, \mbox{\boldmath$x$}, \dot{\mbox{\boldmath$x$}}) + \partial_t f(t, \mbox{\boldmath$x$}) + \dot{\mbox{\boldmath$x$}} \cdot \mbox{\boldmath$\nabla$} f(t, \mbox{\boldmath$x$})
\end{equation}

We can now invoke the gauge principle which prescribes the replacement
\begin{equation}
\partial_t f(t, \mbox{\boldmath$x$}) \rightarrow - q \phi(t, \mbox{\boldmath$x$}) \, , \qquad \mbox{\boldmath$\nabla$} f \rightarrow q \mbox{\boldmath$A$}(t, \mbox{\boldmath$x$}) \, ,
\end{equation}
where $q$ is the coupling parameter (charge). The new function $\phi(t, \mbox{\boldmath$x$})$ and $\mbox{\boldmath$A$}(t, \mbox{\boldmath$x$})$ are the scalar and vector potential of the Maxwell theory. Then the Lagrangian coupled to these potentials reads
\begin{equation}
L^\prime(t, \mbox{\boldmath$x$}, \dot{\mbox{\boldmath$x$}}) = L_0(t, \mbox{\boldmath$x$}, \dot{\mbox{\boldmath$x$}}) - q \phi(t, \mbox{\boldmath$x$}) + q \dot{\mbox{\boldmath$x$}} \cdot \mbox{\boldmath$A$}(t, \mbox{\boldmath$x$})
\end{equation}
With the choice $L_0(t, \mbox{\boldmath$x$}, \dot{\mbox{\boldmath$x$}}) = \frac{1}{2} m  \dot{\mbox{\boldmath$x$}}^2$ we obtain the standard Lorentz force equation of a charged particle moving in an electromagnetic field. 

Now we generalize this approach to higher order Lagrange functions. 

\subsection{Second order formalism}

In the second order formalism we consider Lagrange functions of the form
\begin{equation}
L = L_0(t, \mbox{\boldmath$x$}, \dot{\mbox{\boldmath$x$}}, \ddot{\mbox{\boldmath$x$}})
\end{equation}
from which we obtain the equation of motion
\begin{equation}
0 = \mbox{\boldmath$\nabla$} L - \frac{d}{dt} {\mbox{\boldmath$\nabla$}}_{\dot x} L + \frac{d^2}{dt^2} {\mbox{\boldmath$\nabla$}}_{\ddot x} L = \mbox{\boldmath$\nabla$} L - \frac{d}{dt} \left({\mbox{\boldmath$\nabla$}}_{\dot x} L - \frac{d}{dt} {\mbox{\boldmath$\nabla$}}_{\ddot x} L\right) \, , \label{EoM2}
\end{equation}
where ${\mbox{\boldmath$\nabla$}}_a$ denotes the gradient with respect to the variable $a$. Also in this case we obtain the same equation of motion from another Lagrange function if it differs from the original one by a total time derivative of a function $f$ only. This function, however, now may depend on the velocities $\dot{\mbox{\boldmath$x$}}$
\begin{equation}
L(t, \mbox{\boldmath$x$}, \dot{\mbox{\boldmath$x$}}, \ddot{\mbox{\boldmath$x$}}) \rightarrow L^\prime(t, \mbox{\boldmath$x$}, \dot{\mbox{\boldmath$x$}}, \ddot{\mbox{\boldmath$x$}}) = L_0(t, \mbox{\boldmath$x$}, \dot{\mbox{\boldmath$x$}}, \ddot{\mbox{\boldmath$x$}}) + \frac{d}{dt} f(t, \mbox{\boldmath$x$}, \dot{\mbox{\boldmath$x$}}) \, . 
\end{equation}
The expansion of the total time derivative gives
\begin{equation}
L^\prime(t, \mbox{\boldmath$x$}, \dot{\mbox{\boldmath$x$}}, \ddot{\mbox{\boldmath$x$}}) = L_0(t, \mbox{\boldmath$x$}, \dot{\mbox{\boldmath$x$}}, \ddot{\mbox{\boldmath$x$}}) + \partial_t f(t, \mbox{\boldmath$x$}, \dot{\mbox{\boldmath$x$}}) + \dot{\mbox{\boldmath$x$}} \cdot \mbox{\boldmath$\nabla$} f(t, \mbox{\boldmath$x$}, \dot{\mbox{\boldmath$x$}}) + \ddot{\mbox{\boldmath$x$}} \cdot {\mbox{\boldmath$\nabla$}}_{\dot x} f(t, \mbox{\boldmath$x$}, \dot{\mbox{\boldmath$x$}}) \, . \label{functionf}
\end{equation}

The question now is how to employ the gauge principle. If we replace, e.g., $\partial_t f(t, \mbox{\boldmath$x$}, \dot{\mbox{\boldmath$x$}})$ by a function $\phi(t, \mbox{\boldmath$x$}, \dot{\mbox{\boldmath$x$}})$ then this function cannot describe an external field since it would depend on the velocity. A given external field should be given {\it per se} and should not depend on the state of motion of a  particle. The properties of the external field should be independent of whether the particle is moving through it or not. 

One way to introduce functions depending on time and position only is to assume that the function $f(t, \mbox{\boldmath$x$}, \dot{\mbox{\boldmath$x$}})$ is polynomial in the velocity. That means
\begin{equation}
f(t, \mbox{\boldmath$x$}, \dot{\mbox{\boldmath$x$}}) = \sum_{n = 0}^N f_{i_1 \ldots i_n}(t, \mbox{\boldmath$x$}) \dot x^{i_1} \cdots \dot x^{i_n} \, .
\end{equation}
In this case we can regard the functions $f_{i_1 \ldots i_n}(t, \mbox{\boldmath$x$}) = f_{(i_1 \ldots i_n)}(t, \mbox{\boldmath$x$})$ as gauge functions of an externally given interaction. For this setting we obtain for the new Lagrange function yielding the same equations of motion
\begin{eqnarray}
L^\prime(t, \mbox{\boldmath$x$}, \dot{\mbox{\boldmath$x$}}, \ddot{\mbox{\boldmath$x$}}) & = & L_0(t, \mbox{\boldmath$x$}, \dot{\mbox{\boldmath$x$}}, \ddot{\mbox{\boldmath$x$}}) + \sum_{n = 0}^N \partial_t f_{i_1 \ldots i_n}(t, \mbox{\boldmath$x$}) \dot x^{i_1} \cdots \dot x^{i_n} \nonumber\\
& & + \sum_{n = 0}^N \partial_i f_{i_1 \ldots i_n}(t, \mbox{\boldmath$x$}) \dot x^i \dot x^{i_1} \cdots \dot x^{i_n} + \sum_{n = 1}^N n f_{i_1 \ldots i_n}(t, \mbox{\boldmath$x$}) \ddot x^{i_1} \dot x^{i_2} \cdots \dot x^{i_n} \, . 
\end{eqnarray}
The gauge principle now allows to replace these gauge functions by the gauge fields
\begin{eqnarray}
\partial_t f_{i_1 \ldots i_n}(t, \mbox{\boldmath$x$}) & \rightarrow & - q_n \phi_{i_1 \ldots i_n}(t, \mbox{\boldmath$x$}) \nonumber \\ 
\partial_i f_{i_1 \ldots i_n}(t, \mbox{\boldmath$x$}) & \rightarrow & q_n A_{i i_1 \ldots i_n}(t, \mbox{\boldmath$x$}) \\ 
n f_{i_1 \ldots i_n}(t, \mbox{\boldmath$x$}) & \rightarrow & q_n \psi_{i_1 \ldots i_n}(t, \mbox{\boldmath$x$}) \, ,
\end{eqnarray}
where the $q_n$ are the coupling parameters to these $n^{\rm th}$ rank potentials. The symmetries of these gauge fields are 
\begin{equation}
\phi_{i_1 \ldots i_n}(t, \mbox{\boldmath$x$}) = \phi_{(i_1 \ldots i_n)}(t, \mbox{\boldmath$x$}) \, , \quad A_{i i_1 \ldots i_n}(t, \mbox{\boldmath$x$}) = A_{(i i_1 \ldots i_n)}(t, \mbox{\boldmath$x$}) \, , \quad \psi_{i_1 \ldots i_n}(t, \mbox{\boldmath$x$}) = \psi_{(i_1 \ldots i_n)}(t, \mbox{\boldmath$x$}) \, ,
\end{equation}
that is, all gauge fields are totally symmetric. These fields transform under the generalized gauge transformations as 
\begin{eqnarray}
q_n \phi_{i_1 \ldots i_n}(t, \mbox{\boldmath$x$}) & \rightarrow & q_n \phi_{i_1 \ldots i_n}^\prime(t, \mbox{\boldmath$x$}) = q_n \phi_{i_1 \ldots i_n}(t, \mbox{\boldmath$x$}) - \partial_t f_{i_1 \ldots i_n}(t, \mbox{\boldmath$x$}) \nonumber \\ 
q_n A_{i i_1 \ldots i_n}(t, \mbox{\boldmath$x$}) & \rightarrow & q_n A_{i i_1 \ldots i_n}^\prime(t, \mbox{\boldmath$x$}) = q_n A_{i i_1 \ldots i_n}(t, \mbox{\boldmath$x$}) + \partial_{(i} f_{i_1 \ldots i_n)}(t, \mbox{\boldmath$x$}) \label{GaugeTransformations} \\ 
q_n \psi_{i_1 \ldots i_n}(t, \mbox{\boldmath$x$}) & \rightarrow & q_n \psi_{i_1 \ldots i_n}^\prime(t, \mbox{\boldmath$x$}) = q_n \psi_{i_1 \ldots i_n}(t, \mbox{\boldmath$x$}) + n f_{i_1 \ldots i_n}(t, \mbox{\boldmath$x$})  \, . \nonumber
\end{eqnarray}
For $n = 0$ they reduce to the ordinary Maxwell gauge transformations. 

As a consequence we obtain the Lagrangian of a particle coupled to the new interaction gauge fields
\begin{eqnarray}
L^\prime(t, \mbox{\boldmath$x$}, \dot{\mbox{\boldmath$x$}}, \ddot{\mbox{\boldmath$x$}}) & = & L_0(t, \mbox{\boldmath$x$}, \dot{\mbox{\boldmath$x$}}, \ddot{\mbox{\boldmath$x$}}) - \sum_{n = 0}^N q_n \phi_{i_1 \ldots i_n}(t, \mbox{\boldmath$x$}) \dot x^{i_1} \cdots \dot x^{i_n} \nonumber\\
& & + \sum_{n = 0}^N q_n A_{ii_1 \ldots i_n}(t, \mbox{\boldmath$x$}) \dot x^i \dot x^{i_1} \cdots \dot x^{i_n} + \sum_{n = 0}^N q_n \psi_{i_1 \ldots i_n}(t, \mbox{\boldmath$x$}) \ddot x^{i_1} \dot x^{i_2} \cdots \dot x^{i_n} \, .
\end{eqnarray}
The equations of motion read
\begin{eqnarray}
0 & = & \partial_j L^\prime - \frac{d}{dt} \frac{\partial L^\prime}{\partial \dot x^j} + \frac{d^2}{dt^2} \frac{\partial L^\prime}{\partial \ddot x^j} \nonumber\\
& = & \partial_j L_0 - \frac{d}{dt} \frac{\partial L_0}{\partial \dot x^j} + \frac{d^2}{dt^2} \frac{\partial L_0}{\partial \ddot x^j} - \sum_{n = 0}^N q_n \partial_j \phi_{i_1 \ldots i_n}(t, \mbox{\boldmath$x$}) \dot x^{i_1} \cdots \dot x^{i_n} \nonumber\\
& & + \sum_{n = 0}^N q_n \partial_j A_{ii_1 \ldots i_n}(t, \mbox{\boldmath$x$}) \dot x^i \dot x^{i_1} \cdots \dot x^{i_n} + \sum_{n = 1}^N q_n \partial_j \psi_{i_1 \ldots i_n}(t, \mbox{\boldmath$x$}) \ddot x^{i_1} \dot x^{i_2} \cdots \dot x^{i_n} \nonumber\\ 
& & - \frac{d}{dt} \left(- \sum_{n = 0}^N q_n n \phi_{j i_2 \ldots i_n}(t, \mbox{\boldmath$x$}) \dot x^{i_2} \cdots \dot x^{i_n} + \sum_{n = 0}^N (n+1) q_n A_{ji_1 \ldots i_n}(t, \mbox{\boldmath$x$}) \dot x^{i_1} \cdots \dot x^{i_n} \right. \nonumber\\
& & \left. \qquad\qquad + \sum_{n = 1}^N (n - 1) q_n \psi_{i_1 j i_3 \ldots i_n}(t, \mbox{\boldmath$x$}) \ddot x^{i_1}  \dot x^{i_2} \dot x^{i_3} \cdots \dot x^{i_n}\right)  \nonumber \\ 
& & + \frac{d^2}{dt^2} \left(\sum_{n = 1}^N q_n \psi_{j i_2 \ldots i_n}(t, \mbox{\boldmath$x$}) \dot x^{i_2} \cdots \dot x^{i_n}\right) \, .
\end{eqnarray}
The additional gauge interaction can add a time derivative of at most third order. The principal part of the differential equation remains unaffected. 

Below we typically will use the most simple second order Lagrangian without interaction
\begin{equation}
L_0(t, \mbox{\boldmath$x$}, \dot{\mbox{\boldmath$x$}}, \ddot{\mbox{\boldmath$x$}}) = \frac{m}{2} \dot{\mbox{\boldmath$x$}}^2 - \frac{\epsilon}{2} \ddot{\mbox{\boldmath$x$}}^2 \, . \label{NoetherL0}
\end{equation}
The parameter $\epsilon$ has the dimension ${\rm kg\; s^2}$. If one assumes that this additional term has emerged from influences of quantum gravity then it might be natural to identify it with $\epsilon \sim M_{\rm Pl} T^2_{\rm Pl} \sim 10^{-95} \;{\rm kg \, s^2}$ which is extremely small. 

\section{Noether theorem}

Also for higher order Lagrangians conservation laws can be obtained from the variational principle if we allow non--vanishing variations at the initial and final points. The variations are as usual
\begin{eqnarray}
\bar t & = & t + \tau(t) \label{Variation:t} \\
\bar{\mbox{\boldmath$x$}}(\bar t) & = & \mbox{\boldmath$x$}(t) + \Delta \mbox{\boldmath$x$}(t) \label{Variation:q}\, . 
\end{eqnarray} 
One should bear in mind that here the $\mbox{\boldmath$x$}$ need not to be variables of the configuration space. For these general variations, the variation of the action is
\begin{equation}
\delta S = \bar S - S = \int_{\bar t_1}^{\bar t_2} L(\bar{\mbox{\boldmath$x$}}(\bar t), \dot{\bar{\mbox{\boldmath$x$}}}(\bar t), \ddot{\bar{\mbox{\boldmath$x$}}}(\bar t), \bar t) d\bar t - \int_{t_1}^{t_2} L(\mbox{\boldmath$x$}(t), \dot{\mbox{\boldmath$x$}}(t), \ddot{\mbox{\boldmath$x$}}(t), t) dt \, .
\end{equation}
Proceeding as in the first order case described in textbooks we arrive at  
\begin{eqnarray}
\delta S & = & \int_{t_1}^{t_2} \left(\mbox{\boldmath$\nabla$} L - \frac{d}{dt} {\mbox{\boldmath$\nabla$}}_{\dot x} L + \frac{d^2}{dt^2} {\mbox{\boldmath$\nabla$}}_{\ddot x} L\right) \delta\mbox{\boldmath$x$} dt \nonumber\\
& & + \left({\mbox{\boldmath$\nabla$}}_{\ddot x} L \cdot \frac{d}{dt} \left(\Delta\mbox{\boldmath$x$} - \tau \dot{\mbox{\boldmath$x$}}\right) + \left({\mbox{\boldmath$\nabla$}}_{\dot x} L - \frac{d}{dt} {\mbox{\boldmath$\nabla$}}_{\ddot x} L\right) \cdot \left(\Delta\mbox{\boldmath$x$} - \tau \dot{\mbox{\boldmath$x$}}\right) + \tau L\right)_{t_1}^{t_2} \, .
\end{eqnarray}
If the equations of motion are fulfilled, and if the action is invariant under the variations (\ref{Variation:t},\ref{Variation:q}), then we obtain the conserved quantity 
\begin{equation}
{\mbox{\boldmath$p$}}_2 \cdot \frac{d}{dt} \left(\Delta\mbox{\boldmath$x$} - \tau \dot{\mbox{\boldmath$x$}}\right) + \left({\mbox{\boldmath$p$}}_1 - \dot{\mbox{\boldmath$p$}}_2\right) \cdot \left(\Delta\mbox{\boldmath$x$} - \tau \dot{\mbox{\boldmath$x$}}\right) + \tau L = const \, , \label{Variation:Noether0}
\end{equation}
where we defined the momenta
\begin{equation}
{\mbox{\boldmath$p$}}_1 = {\mbox{\boldmath$\nabla$}}_{\dot x} L \, , \qquad \text{and} \qquad {\mbox{\boldmath$p$}}_2 = {\mbox{\boldmath$\nabla$}}_{\ddot x} L \, .
\end{equation}

If the action does not vanish but, instead, changes with a total time derivative of a function $F(\mbox{\boldmath$x$}(t), \dot{\mbox{\boldmath$x$}}(t), t)$, then the equations of motion in terms of the Euler--Lagrange equations do not change and we  have a modified conserved quantity
\begin{equation}
{\mbox{\boldmath$p$}}_2 \cdot \frac{d}{dt} \left(\Delta\mbox{\boldmath$x$} - \tau \dot{\mbox{\boldmath$x$}}\right) + \left({\mbox{\boldmath$p$}}_1 - \dot{\mbox{\boldmath$p$}}_2\right) \cdot \left(\Delta\mbox{\boldmath$x$} - \tau \dot{\mbox{\boldmath$x$}}\right) + \tau L + F = const. \label{Variation:ConservedQuantityTotalTimeDerivative}
\end{equation}
From this general Noether theorem we derive the following conserved quantities:

\subsection{Momentum conservation}

At first we consider the transformations $\tau(t) = 0$ and $\Delta \mbox{\boldmath$x$}(t) = const$. We obtain the conserved momentum
\begin{equation}
\mbox{\boldmath$P$} = {\mbox{\boldmath$p$}}_1 - \dot{\mbox{\boldmath$p$}}_2 = const.
\end{equation}
This may also directly be inferred from the Euler--Lagrange equations of motion \eqref{EoM2}. For the Lagrangian \eqref{NoetherL0} we get
\begin{equation}
\mbox{\boldmath$P$} = m \dot{\mbox{\boldmath$x$}} + \epsilon \dddot{\mbox{\boldmath$x$}} = const. 
\end{equation}

\subsection{Energy conservation}

Next we consider the transformations $\tau(t) = \tau_0$ and $\Delta \mbox{\boldmath$x$}(t) = 0$. The corresponding conserved energy is
\begin{equation}
E = {\mbox{\boldmath$p$}}_2 \cdot \ddot{\mbox{\boldmath$x$}} + \left({\mbox{\boldmath$p$}}_1 - \dot{\mbox{\boldmath$p$}}_2\right) \cdot \dot{\mbox{\boldmath$x$}} - L = {\mbox{\boldmath$p$}}_2 \cdot \ddot{\mbox{\boldmath$x$}} + \mbox{\boldmath$P$} \cdot \dot{\mbox{\boldmath$x$}} - L = const \, . \label{ConservedEnergy}
\end{equation}
For the Lagrangian (\ref{NoetherL0}) we obtain 
\begin{equation}
E = \frac{1}{2} m \dot{\mbox{\boldmath$x$}}^2 + \frac{1}{2} \epsilon \left(2 \dddot{\mbox{\boldmath$x$}} \cdot \dot{\mbox{\boldmath$x$}}-\ddot{\mbox{\boldmath$x$}}^2\right) \, .
\end{equation} 

\subsection{Angular momentum conservation}

In the next example we assume that the action is invariant under the transformations
$\tau(t) = 0$ and $\Delta \mbox{\boldmath$x$} = \Delta\mbox{\boldmath$\varphi$} \times \mbox{\boldmath$x$}$. This corresponds to a rotation and the corresponding conserved angular momentum is 
\begin{equation}
\mbox{\boldmath$L$} = \mbox{\boldmath$x$} \times \left({\mbox{\boldmath$p$}}_1 - \dot{\mbox{\boldmath$p$}}_2\right) + \dot{\mbox{\boldmath$x$}} \times {\mbox{\boldmath$p$}}_2 = \mbox{\boldmath$x$} \times \mbox{\boldmath$P$} + \dot{\mbox{\boldmath$x$}} \times {\mbox{\boldmath$p$}}_2 = const.
\end{equation}
For the Lagrangian (\ref{NoetherL0}) we obtain the conserved angular momentum
\begin{equation}
\mbox{\boldmath$L$} = \mbox{\boldmath$x$} \times (m \dot{\mbox{\boldmath$x$}}) + \epsilon \left( \mbox{\boldmath$x$} \times \dddot{\mbox{\boldmath$x$}} - \dot{\mbox{\boldmath$x$}} \times \ddot{\mbox{\boldmath$x$}}\right) = \mbox{\boldmath$x$} \times \mbox{\boldmath$P$} - \epsilon \dot{\mbox{\boldmath$x$}} \times \ddot{\mbox{\boldmath$x$}} = const.
\end{equation}

\subsection{Proper Galileo transformation}

At last we consider the Galilei transformations $\tau(t) = 0$ and $\Delta{\mbox{\boldmath$x$}} = \Delta \mbox{\boldmath$v$} t$ for the Lagrangian \eqref{NoetherL0}, 
where $\Delta \mbox{\boldmath$v$}$ is assumed to be very small. The term $\frac{1}{2} m \dot{\mbox{\boldmath$x$}}^2$ changes by a total differential of $m \mbox{\boldmath$x$} \cdot \Delta\mbox{\boldmath$v$}$ so that we obtain the conserved quantity
\begin{equation}
C = {\mbox{\boldmath$p$}}_2 + \left({\mbox{\boldmath$p$}}_1 - \dot{\mbox{\boldmath$p$}}_2\right) t - m \mbox{\boldmath$x$} 
\end{equation}
from which we deduce a uniform motion
\begin{equation}
\mbox{\boldmath$x$} =\frac{{\mbox{\boldmath$p$}}_2}{m} + \frac{{\mbox{\boldmath$p$}}_1 - \dot{\mbox{\boldmath$p$}}_2}{m} t + {\mbox{\boldmath$x$}}_0 \, . 
\end{equation}

\section{The most simple gauge model, $n = 0$}

\subsection{Equation of motion}

The most simple case with non--trivial dynamics is given for the special case that $f$ in \eqref{functionf} is a function of $t$ and $\mbox{\boldmath$x$}$ only. Then we have
\begin{equation}
L^\prime(t, \mbox{\boldmath$x$}, \dot{\mbox{\boldmath$x$}}, \ddot{\mbox{\boldmath$x$}}) = L_0(t, \mbox{\boldmath$x$}, \dot{\mbox{\boldmath$x$}}, \ddot{\mbox{\boldmath$x$}}) - q \phi(t, \mbox{\boldmath$x$}) + q \dot x^i A_i(t, \mbox{\boldmath$x$}) \, .
\end{equation}
For $L_0$ from \eqref{NoetherL0} the equations of motion read
\begin{equation}
\epsilon \stackrel{....}{\mbox{\boldmath$x$}} + m \ddot{\mbox{\boldmath$x$}} = q \mbox{\boldmath$E$}(t, \mbox{\boldmath$x$}) + q \dot{\mbox{\boldmath$x$}} \times {\mbox{\boldmath$B$}}(t, \mbox{\boldmath$x$}) \, , \label{4eomMaxwell}
\end{equation}
where ${\mbox{\boldmath$E$}}$ and ${\mbox{\boldmath$B$}}$ are the electric and magnetic field derived as usual from the scalar and vector potentials $\phi$ and $\mbox{\boldmath$A$}$. This equation of motion is the standard one with a small additional forth order term. 

We may simplify even further by taking a vanishing magnetic field, $\mbox{\boldmath$B$} = 0$ and a constant electric field ${\mbox{\boldmath$E$}}(\mbox{\boldmath$x$}) = {\mbox{\boldmath$E$}}_0 = const$,
\begin{equation}
\epsilon \stackrel{....}{\mbox{\boldmath$x$}} + m \ddot{\mbox{\boldmath$x$}} = q {\mbox{\boldmath$E$}}_0 \, .
\end{equation}
This is the equation we like to solve now. 

\subsection{Particle motion}

The first two time integrations are easily performed and give
\begin{equation}
\epsilon \ddot{\mbox{\boldmath$x$}} + m \mbox{\boldmath$x$} = \frac{q}{2} {\mbox{\boldmath$E$}}_0 (t - t_0)^2 + \mbox{\boldmath$a$} (t - t_0) + \mbox{\boldmath$b$} \, , 
\end{equation}
where $\mbox{\boldmath$a$}$ and $\mbox{\boldmath$b$}$ are integration constants. Next we introduce a new function $\bar{\mbox{\boldmath$x$}}$ through $\mbox{\boldmath$x$} = {\mbox{\boldmath$x$}}_0 + \epsilon \bar{\mbox{\boldmath$x$}}$ with ${\mbox{\boldmath$x$}}_0 = \frac{q}{2 m} {\mbox{\boldmath$E$}}_0 (t - t_0)^2 + \frac{1}{m} \mbox{\boldmath$a$} (t - t_0) + \frac{1}{m}  \mbox{\boldmath$b$}$ which represents the solution of the corresponding equation of motion without the forth order term. Since $\epsilon$ will be assumed to be small, the deviation from the standard solution should be small, too. Therefore we introduced an extra $\epsilon$ in front of $\bar{\mbox{\boldmath$x$}}$. The differential equation for $\bar{\mbox{\boldmath$x$}}$ then reads
\begin{equation}
\ddot{\bar{\mbox{\boldmath$x$}}} + \frac{m}{\epsilon} \bar{\mbox{\boldmath$x$}} = - \frac{q}{m \epsilon} {\mbox{\boldmath$E$}}_0 \, .
\end{equation}
A further substitution $\hat{\mbox{\boldmath$x$}} = \bar{\mbox{\boldmath$x$}} + \frac{\epsilon}{m} \frac{q}{m \epsilon} {\mbox{\boldmath$E$}}_0$ yields the equation for a particle in a harmonic potential
\begin{equation}
\ddot{\hat{\mbox{\boldmath$x$}}} + \frac{m}{\epsilon} \hat{\mbox{\boldmath$x$}} = 0 \, ,
\end{equation}
which, according to the sign of $\epsilon$, possess the solution
\begin{align}
\hat{\mbox{\boldmath$x$}} & = {\mbox{\boldmath$A$}} \cos\left(\omega t\right) + {\mbox{\boldmath$B$}} \sin\left(\omega t\right)\qquad & \text{for} \quad \epsilon > 0 \\ 
\hat{\mbox{\boldmath$x$}} & = {\mbox{\boldmath$A$}}_1 \cosh\left(\omega t\right) + {\mbox{\boldmath$A$}}_2 \sinh\left(\omega t\right)\qquad & \text{for} \quad \epsilon < 0
\end{align}
for some amplitudes $\mbox{\boldmath$A$}$, $\mbox{\boldmath$B$}$, ${\mbox{\boldmath$A$}}_1$, and ${\mbox{\boldmath$A$}}_2$ and with $\omega = \sqrt{m/\epsilon}$. 

As a consequence we arrived for $\epsilon > 0$ at the solution 
\begin{equation}
\mbox{\boldmath$x$}(t) = \frac{q}{2 m} {\mbox{\boldmath$E$}}_0 (t - t_0)^2 + \frac{1}{m} \mbox{\boldmath$a$} (t - t_0) + \frac{1}{m} \mbox{\boldmath$b$} + \epsilon \mbox{\boldmath$A$} \cos\left(\omega t\right) + \epsilon \mbox{\boldmath$B$} \sin\left(\omega t\right) - \frac{\epsilon}{m} \frac{q}{m} {\mbox{\boldmath$E$}}_0 \, . \label{Solution1stModel}
\end{equation}
In the limit $\epsilon \rightarrow 0$ we obtain the standard solution. The velocity is 
\begin{equation}
\dot{\mbox{\boldmath$x$}}(t) = \frac{q}{m} {\mbox{\boldmath$E$}}_0 (t - t_0) + \frac{1}{m} \mbox{\boldmath$a$} - \sqrt{m \epsilon} \mbox{\boldmath$A$} \sin\left(\omega t\right) + \sqrt{m \epsilon} \mbox{\boldmath$B$} \cos\left(\omega t\right) \, ,
\end{equation}
which also approaches the standard expression for $\epsilon \rightarrow 0$. The acceleration
\begin{equation}
\ddot{\mbox{\boldmath$x$}}(t) = \frac{q}{m} {\mbox{\boldmath$E$}}_0 - m \mbox{\boldmath$A$} \cos\left(\omega t\right)- m \mbox{\boldmath$B$} \sin\left(\omega t\right) \, ,
\end{equation}
however, has a large fluctuating term of order 1. 

For very small positive $\epsilon$ the additional term in the path \eqref{Solution1stModel} is very small but fast oscillating {\it zitterbewegung}. One may turn the above result into a positive statement: At least within a Lagrangian approach and assuming that higher order terms are small, the paths originating from a corresponding higher order modification are rather inert against these modifications. The mean orbits behave like orbits given by second order equations of motion. We will present some ways to experimentally search for these fundamental oscillations in Section \ref{Experiemtn}. 

This result may be extended to the case of slowly varying arbitrary electromagnetic fields. The equation of motion for an arbitrary electromagnetic field \eqref{4eomMaxwell} may be attacked through the substitution $\mbox{\boldmath$x$} = \epsilon \bar{\mbox{\boldmath$x$}} + {\mbox{\boldmath$x$}}_0$ where ${\mbox{\boldmath$x$}}_0$ is assumed to solve the equation of motion without the forth order term. If we assume that the force $\mbox{\boldmath$F$}(\mbox{\boldmath$x$}) = q \mbox{\boldmath$E$}(\mbox{\boldmath$x$}) + q \mbox{\boldmath$v$} \times \mbox{\boldmath$B$}(\mbox{\boldmath$x$})$ is very smooth and that the deviation $\epsilon \bar{\mbox{\boldmath$x$}}$ is very small (that is, if $\bar{\mbox{\boldmath$x$}} \cdot \mbox{\boldmath$\nabla$} \mbox{\boldmath$F$} \ll m \bar{\mbox{\boldmath$x$}}$ and can be neglected), then we obtain 
\begin{equation}
\stackrel{....}{{\mbox{\boldmath$x$}}_0} + \epsilon \stackrel{....}{\bar{\mbox{\boldmath$x$}}} + m \ddot{\bar{\mbox{\boldmath$x$}}} = 0 \, .
\end{equation}
This can be integrated twice
\begin{equation}
\ddot{\mbox{\boldmath$x$}}_0 + \epsilon \ddot{\bar{\mbox{\boldmath$x$}}} + m \bar{\mbox{\boldmath$x$}} = \mbox{\boldmath$a$} t + \mbox{\boldmath$b$} \, , 
\end{equation}
where $\mbox{\boldmath$a$}$ and $\mbox{\boldmath$b$}$ are two integration constants. Inserting the equation for $\ddot{\mbox{\boldmath$x$}}_0$ yields
\begin{equation}
\ddot{\bar{\mbox{\boldmath$x$}}} + \frac{m}{\epsilon} \bar{\mbox{\boldmath$x$}} = - \frac{1}{m \epsilon} \mbox{\boldmath$F$}({\mbox{\boldmath$x$}}_0) + \frac{1}{\epsilon} \mbox{\boldmath$a$} t + \frac{1}{\epsilon} \mbox{\boldmath$b$} \, .
\end{equation}
With a new variable $\hat{\mbox{\boldmath$x$}} = \bar{\mbox{\boldmath$x$}} - \frac{1}{m} \mbox{\boldmath$a$} t + \frac{1}{m} \mbox{\boldmath$b$} - \frac{1}{m^2} \mbox{\boldmath$F$}({\mbox{\boldmath$x$}}_0)$ we have
\begin{equation}
\ddot{\hat{\mbox{\boldmath$x$}}} + \frac{m}{\epsilon} \hat{\mbox{\boldmath$x$}} = 0 \, .
\end{equation}
Then again, for positive $\epsilon$, $\hat{\mbox{\boldmath$x$}}$ is a fast oscillating term which adds to the standard solution. The total solution then is
\begin{equation}
\mbox{\boldmath$x$}(t) = {\mbox{\boldmath$x$}}_0(t) + \epsilon \left(\hat{\mbox{\boldmath$x$}}(t) + \frac{1}{m} \mbox{\boldmath$a$} t - \frac{1}{m} \mbox{\boldmath$b$} + \frac{1}{m^2} \mbox{\boldmath$F$}({\mbox{\boldmath$x$}}_0(t))\right) \, .
\end{equation}
This solution consists of the standard solution ${\mbox{\boldmath$x$}}_0(t)$ which is the main motion, a small position--dependent displacement and a small linearly growing term, and a small fast oscillating term, a kind of {\it zitterbewegung}. From ordinary observations, $\mbox{\boldmath$a$}$ and $\mbox{\boldmath$b$}$ should be very small. Neglecting these particular solutions, then the standard solution of the standard second order equation of motion seems to be rather robust against adding a higher order term. 

%For small masses this approximation scheme breaks down. In the case of gravity, the force $\mbox{\boldmath$F$} = m \mbox{\boldmath$\nabla$} U$, where $U$ is the Newtonian potential. Then we have
%\begin{equation}
%\ddot{\hat{\mbox{\boldmath$x$}}} + \frac{m}{\epsilon} \hat{\mbox{\boldmath$x$}} = - \frac{1}{\epsilon} \mbox{\boldmath$\nabla$}U({\mbox{\boldmath$x$}}_0) \, .
%\end{equation}
%Then in the limit of large masses, the reference to the gravitational field disappears thus making the {\it zitterbewegung} a universal effect which is independent of the gravitational field. 

\subsection{Conserved energy}

The conserved energy in this most simple model reads
\begin{equation}
E = \frac{m}{2} \dot{\mbox{\boldmath$x$}}^2 + \frac{\epsilon}{2} \left(2 \dddot{\mbox{\boldmath$x$}} \cdot \dot{\mbox{\boldmath$x$}} - \ddot{\mbox{\boldmath$x$}}^2\right) + q \phi \, ,
\end{equation}
which also can be obtained by multiplying the equation of motion by $\dot{\mbox{\boldmath$x$}}$ and partial integration. 

\section{The next gauge model, $n = 1$}

\subsection{The equation of motion}

Here we now like to explore the model where $f$ is a function which is polynomial of first order in the velocities, that is, $f(t, \mbox{\boldmath$x$}, \dot{\mbox{\boldmath$x$}}) = f^{(0)}(t, \mbox{\boldmath$x$}) + f^{(1)}_i(t, \mbox{\boldmath$x$}) \dot x^i$. Then the corresponding gauged Lagrange function reads
\begin{equation}
L(t, \mbox{\boldmath$x$}, \dot{\mbox{\boldmath$x$}}, \ddot{\mbox{\boldmath$x$}}) = L_0(t, \mbox{\boldmath$x$}, \dot{\mbox{\boldmath$x$}}, \ddot{\mbox{\boldmath$x$}}) - q_0 \phi + q_0 \dot x^i A_i - q_1 \dot x^i \phi_i + q_1 \dot x^i \dot x^j A_{ij} + q_1 \ddot x^i \psi_i \, , \label{n1gaugesLagrangian}
\end{equation}
where all functions depend on $t$ and $\mbox{\boldmath$x$}$. Here $q_0$ is the coupling related to $f^{(0)}$, and $q_1$ the coupling related to $f_i$. Beside the usual scalar and vector potential, $\phi$ and $A_i$ we have in addition two vector potentials $\phi_i$ and $\psi_i$ and a tensorial potential $A_{ij}$. The gauge transformations are given by
\begin{equation}
\begin{split}
q_0 \phi^\prime &= q_0 \phi - \partial_t f^{(0)} \, , \qquad q_0 A_i^\prime = q_0 A_i + \partial_i f^{(0)} \\ 
q_1 \phi_i^\prime = q_1 \phi - \partial_t f_i^{(1)} \, , & \qquad q_1 \psi_i^\prime = q_1 \psi_i + f_i^{(1)} \, , \qquad q_1 A_{ij}^\prime = q_1 A_{ij} + \partial_{(i} f_{i)}^{(1)} \, . 
\end{split} \label{gaugetrans2}
\end{equation}

The corresponding Euler--Lagrange equation is
\begin{eqnarray}
0 & = & \frac{\partial L_0}{\partial x^i} - \frac{d}{dt} \frac{\partial L_0}{\partial \dot x^i} + \frac{d^2}{dt^2} \frac{\partial L_0}{\partial \ddot x^i} \label{1line} \\
& & - q_0 \partial_i \phi - q_0 \partial_t A_i + q_1 \partial_t \phi_i + q_1 \partial_t^2 \psi_i  \label{2line}\\
& & + \dot x^j \left(\partial_i (q_0 A_j - q_1 \phi_j) - \left(\partial_j q_0 A_i - q_1 \partial_j \phi_i\right) + 2 q_1 \partial_t \partial_{[j} \psi_{i]}\right)  \label{3line} \\
& & - 2 q_1 \left(\ddot x^j \tilde A_{ij} + \frac{1}{2} (\partial_k \tilde A_{ij} + \partial_j \tilde A_{ik} - \partial_i \tilde A_{jk}) \dot x^j \dot x^k\right) - 2 q_1 \partial_t \tilde A_{ij} \dot x^j  \, , \label{4line}
\end{eqnarray}
where we defined $\tilde A_{ij} := A_{ij} - \partial_{(i} \psi_{j)}$. 
It should be noted that the gauge terms do not produce any $\dddot x$--terms. 

The first line \eqref{1line} is the (still unspecified) equation of motion of the free particle, the second line \eqref{2line} describes a force due to two electric fields ${\mbox{\boldmath$E$}}^{(0)} := - \mbox{\boldmath$\nabla$} \phi - \partial_t \mbox{\boldmath$A$}$ and ${\mbox{\boldmath$E$}}^{(1)} := \partial_t \mbox{\boldmath$\phi$} + \partial_t^2 \mbox{\boldmath$\psi$}$, the third line \eqref{3line} is a Lorentz--like force with the two magnetic fields ${\mbox{\boldmath$B$}}^{(0)} := \mbox{\boldmath$\nabla$} \times \mbox{\boldmath$A$}$ and ${\mbox{\boldmath$B$}}^{(1)} := \mbox{\boldmath$\nabla$} \times (-\mbox{\boldmath$\phi$} + \partial_t \mbox{\boldmath$\psi$})$, which are all gauge invariant. The forth line \eqref{4line} resembles the form of covariant derivative with a connection based on a second rank tensor $\tilde A_{ij}$.  %Assuming that the tensor $\tilde A_{ij}$ ist not singular then we can write
%\begin{equation}
%0 = \frac{\partial L_0}{\partial x^i} - \frac{d}{dt} \frac{\partial L_0}{\partial \dot x^i} + \frac{d^2}{dt^2} \frac{\partial L_0}{\partial \ddot x^i} + q_0 {\mbox{\boldmath$E$}}^{(0)} + q_1 {\mbox{\boldmath$E$}}^{(1)} + \dot{\mbox{\boldmath$x$}} \times (q_0 {\mbox{\boldmath$B$}}^{(0)} + q_1 {\mbox{\boldmath$B$}}^{(1)}) - 2  q_1\tilde A_{ij} (\tilde D_{\dot x} \dot x)^j - 2 q_1 \partial_t \tilde A_{ij} \dot x^j
%\end{equation}
%where $\tilde D_{\dot x}$ is the covariant derivative along $\dot x$ where the Christoffel symbols are based on $\tilde A_{ij}$. 

Taking again $L_0$ of the form \eqref{NoetherL0} yields an equation of motion of forth order of the form
\begin{equation}
\epsilon \stackrel{....}{x}^j \delta_{ij} + m g_{ij} (D_{\dot x} \dot x)^j - m \partial_t g_{ij} \dot x^j = q_0 E_i^{(0)} + q_1 E_i^{(1)} + (\dot{\mbox{\boldmath$x$}} \times (q_0 {\mbox{\boldmath$B$}}^{(0)} + q_1 {\mbox{\boldmath$B$}}^{(1)}))_i \, . \label{finaleom}
\end{equation} 
where we introduced an effective 3--metric
\begin{equation}
g_{ij} = \delta_{ij} + 2 \frac{q_1}{m} \tilde A_{ij} \label{metric}
\end{equation}
and where the covariant derivative $D_{\dot x}$ is formulated with a Christoffel symbol based on $g_{ij}$. This metric is invariant under the gauge transformations \eqref{gaugetrans2}. The second and third term on the left hand side can be regarded as equation of motion arising from the variation of $\int g_{ij}(t, \mbox{\boldmath$x$}) \dot x^i \dot x^j dt$. If in \eqref{finaleom} we let $\epsilon \rightarrow 0$ then only the forth order derivative term will vanish, the other terms, in particular the metric, remain. 

One should note that by means of our second order gauge principle we were able to establish a space--metric as an ordinary gauge field. Together with this metric gauge field we also introduced an interaction with an electromagnetic gauge field. Therefore, by means of {\it one} gauge procedure we introduced the the interaction with a metric as well as with an electromagnetic field. 

By omitting the $n = 0$ part of this gauge formalism, or equivalently, by setting $q_0 = 0$, we have a combined gauge formalism for the metric and an electromagnetic field. We also can choose ${\mbox{\boldmath$E$}}^{(1)} = 0$ and ${\mbox{\boldmath$B$}}^{(1)} = 0$ while keeping $\tilde A_{ij} \neq 0$, ${\mbox{\boldmath$E$}}^{(0)} \neq 0$, and ${\mbox{\boldmath$B$}}^{(0)} \neq 0$. This latter choice introduces the electromagnetic field and the space metric independently. 

The structure of the above equation of motion \eqref{finaleom} is 
\begin{equation}
\epsilon \stackrel{....}{\mbox{\boldmath$x$}} + m \ddot{\mbox{\boldmath$x$}} = \mbox{\boldmath$K$}(t, \mbox{\boldmath$x$}, \dot{\mbox{\boldmath$x$}}) \, ,
\end{equation}
where $\mbox{\boldmath$K$}(t, \mbox{\boldmath$x$}, \dot{\mbox{\boldmath$x$}})$ is a polynomial of order two in the velocities what slightly generalizes \eqref{4eomMaxwell}. 
As a consequence, for ordinary situations (smooth forces) and small $\epsilon$ we again obtain some {\it zitterbewegung} resulting from the inclusion of higher order derivatives. Again, the main (mean) motion is described by the second order part of the equation of motion. 

\subsection{The conserved energy}

We assume that all fields do not depend on time and determine the conserved energy. Then again we obtain an expression where the space metric appears at the right place. Applying \eqref{ConservedEnergy} yields
\begin{equation}
E = \frac{1}{2} m g_{ij} \dot x^i \dot x^j - q_0 \phi + \frac{1}{2} \epsilon \left(\ddot{\mbox{\boldmath$x$}}^2 - 2 \dddot{\mbox{\boldmath$x$}} \cdot \dot{\mbox{\boldmath$x$}}\right) \, .
\end{equation}
As expected, the effective metric \eqref{metric} enters the kinetic energy. 

\subsection{Relativistic formulation}

It is clear from $\ddot x^i \psi_i = - \dot x^i \frac{d}{dt} \psi_i = - \dot x^i \partial_t \psi_i - \dot x^j \dot x^j \partial_j \psi_i$ which holds modulo a total time derivative, that 
\begin{equation}
- q_1 \dot x^i \phi_i + q_1 \dot x^i \dot x^j A_{ij} + q_1 \ddot x^i \psi_i = - q_1 \dot x^i \left(\phi_i + \partial_t \psi_i\right) + q_1 \dot x^i \dot x^j \left(A_{ij} - \partial_j \psi_i\right) \, ,
\end{equation}
what explains the definitions of ${\mbox{\boldmath$E$}}^{(1)}$ and ${\mbox{\boldmath$B$}}^{(1)}$. Furthermore, with $x^a = (t, x^i)$, $a = 0, \ldots, 3$, and the 3+1 splitting 
\begin{equation}
q_0 A_a \dot x^a + q_1 A_{ab} \dot x^a \dot x^b = q_0 A_0 + q_1 A_{00} + \left(q_0 A_i + 2 q_1 A_{i0}\right) \dot x^i + q_1 A_{ij} \dot x^i \dot x^j \, ,
\end{equation}
we can reproduce, with redefinitions, the gauge ansatz \eqref{n1gaugesLagrangian} above. Therefore, the gauge field part in the Lagrangian \eqref{n1gaugesLagrangian} can be written in 4--covariant form
\begin{equation}
L(t, \mbox{\boldmath$x$}, \dot{\mbox{\boldmath$x$}}, \ddot{\mbox{\boldmath$x$}}) = L_0(t, \mbox{\boldmath$x$}, \dot{\mbox{\boldmath$x$}}, \ddot{\mbox{\boldmath$x$}}) - q_0 A_a \dot x^a + q_1 A_{ab} \dot x^a \dot x^b \, . \label{n1gaugesLagrangiancov}
\end{equation}
It is the last term quadratic in the velocities which, when interpreted as gauge field, requires the second order Lagrangian. 

%@@@   $L_0(x^{\mu},u^{\mu}, a^{\mu}, \tau) = - mc \sqrt{u^{\mu}u_{\mu}} - \epsilon a^{\mu}a_{\mu}$   @@@

\section{Comparison with Experiments}\label{Experiemtn}

We like to describe possible experiments which may be sensitive to the higher order modifications in the equations of motion. We take situations where a charged particle is placed in a constant electric field, e.g., within a condenser. In standard theory the charged particle will be accelerated and the final velocity depends on the voltage only, not on the spacing between the plates of the condenser. To make the situation as simple as possible we consider a one--dimensional problem and take as initial conditions that the particle is at absolute rest at $t_0 = 0$. 

We take our general solution \eqref{Solution1stModel} (in one dimension and with $t_0 = 0$) 
\begin{eqnarray}
x(t) & = & \frac{q}{2 m} E_0 t^2 + \frac{1}{m} a t + \frac{1}{m} b + \epsilon A \cos\left(\omega t\right) + \epsilon B \sin\left(\omega t\right) - \frac{\epsilon}{m} \frac{q}{m} E_0 \\
\dot x(t) & = & \frac{q}{m} E_0 t + \frac{1}{m} a - A \sqrt{m \epsilon} \sin\left(\omega t\right) + B \sqrt{m \epsilon} \cos\left(\omega t\right) \\
\ddot x(t) & = & \frac{q}{m} E_0 - A m \cos\left(\omega t\right) - B m \sin\left(\omega t\right) \\
\dddot x(t) & = & \epsilon A \left(\frac{m}{\epsilon}\right)^{\frac32} \sin\left(\omega t\right) - \epsilon B \left(\frac{m}{\epsilon}\right)^{\frac32} \cos\left(\omega t\right) 
\end{eqnarray}
and determine the parameters in terms of the initial conditions $x(0) = 0$, $\dot x(0) = 0$, $\ddot x(0) = 0$, and $\dddot x(0) = 0$. We obtain $B = 0$, $a = 0$, $A = \frac{q}{m^2} E_0$, and $b = 0$. The solution then reads
\begin{equation}
x(t) = \frac{q}{m} E_0 \left(\frac{1}{2} t^2 + \frac{\epsilon}{m} \left( \cos\left(\omega t\right) - 1\right)\right)
\end{equation}
and we also have
\begin{eqnarray}
\dot x(t) & = & \frac{q}{m} E_0 \left(t - \sqrt{\frac{\epsilon}{m}} \sin\left(\omega t\right)\right)  \\
\ddot x(t) & = & \frac{q}{m} E_0 \left(1 - \cos\left(\omega t\right)\right) \label{anomalousacceleration} \\
\dddot x(t) & = & \frac{q}{m} E_0 \sqrt{\frac{m}{\epsilon}} \sin\left(\omega t\right) \, .
\end{eqnarray}

Now we discuss three different measurements which outcome depend on $\epsilon$: (i) the time of flight in an accelerator, (ii) position and velocity variance, (iii) noise in electronic circuits. We always calculate first order effects. We also do not take into account possible effects of radiation reaction. Due to the modified fundamental equation of motion this needs a new derivation of this effect which will be planned for the future. 

\subsection{Determination of time of flight}

In the standard theory ($\epsilon = 0$ in the Lagrangian) we have the solution  $x(t) = \frac{q}{2 m} E_0 t^2$. For a charged particle being accelerated through its motion through an accelerator of length $L$ we have $L = \frac{q}{2 m} \frac{\Delta\phi}{L} t^2$, where $\Delta\phi$ is the potential difference the particle has to traverse. Then $t^2 = \frac{2 m}{q \Delta\phi} L^2$ so that $t$ is proportional to the spacing $L$ and the final velocity does not depend on the distance $L$. This will be different for our higher order theory. 

We determine the time the particle needs to traverse the length $L$. For that we first have to calculate $t$ from $L = x(t)$:
\begin{equation}
L = \frac{q}{m} E_0 \left(\frac{1}{2} t^2 + \frac{\epsilon}{m} \left( \cos\left(\omega t\right) - 1\right)\right) \, ,
\end{equation}
what is a transcendental equation. For $\epsilon = 0$ we have $t = t_0 = \sqrt{\frac{2 m L}{q E_0}}$. Therefore we make the approximative ansatz $t = t_0 + t_1$, where $t_1 \ll t_0$ and $t_1$ is of the order $\epsilon$. Solving for $t_1$ gives the time of flight to first order correction in $\epsilon$ so that
\begin{equation}
t = \sqrt{\frac{2 m L}{q E_0}} - \sqrt{\frac{q E_0}{2 m L}} \frac{\epsilon}{m} \left(\cos\left(\omega \sqrt{\frac{2 m L}{q E_0}}\right) - 1\right)
\end{equation}
We use this time in order to calculate the velocity at $x(t) = L$ and obtain 
\begin{eqnarray}
\dot x(L)  =  \dot x_0 \left(1 + \frac{\epsilon}{4 m} \frac{\dot x_0^2}{L^2} \left(1-\cos\left(\omega \sqrt{- \frac{m L^2}{2 q \Delta\phi}}\right)\right) + \sqrt{\frac{\epsilon}{4 m}} \frac{\dot x_0}{L}  \sin\left(\omega \sqrt{- \frac{2 m L^2}{q \Delta\phi}}\right)\right) \, , 
\end{eqnarray}
where we substituted $E_0 = - \frac{\Delta\phi}{L}$ and also inserted the velocity of the standard theory $\dot x_0 = \sqrt{- \frac{2 q \Delta\phi}{m}}$. In the standard theory $\dot x(L)$ does not depend on $L$. Here, by varying $L$ we get first oscillations in the velocity due to the sin and cos terms, but also an offset $\frac{\epsilon}{4m} \frac{\dot x_0^3}{L^2}$. 

Since $\epsilon$ is assumed to be small, then a change in $L$ will result in fast oscillations which probably cannot be resolved. Therefore, averaging over a small $L$ interval yields
\begin{equation}
\langle \dot x(L) \rangle = \dot x_0 \left(1 + \frac{\epsilon}{4 m} \frac{\dot x_0^2}{L^2}\right) \, .
\end{equation}
Therefore in the mean the velocity after traversing the condenser is a bit larger. 

Rewriting the above result as relative velocity deviation
\begin{equation}
\frac{\langle \dot x(L) \rangle - \dot x_0}{\dot x_0} = \frac{\epsilon}{4 m} \frac{\dot x_0^2}{L^2}
\end{equation}
it is clear that one obtains good estimates for $\epsilon$ for large velocities $\dot x_0$, short $L$ and small $m$. Taking, e.g., an ion of mass of $m = 60\;{\rm u}$, a final energy of 10 MeV, a traversed distance of $L = 1\;{\rm m}$, and an accuracy to measure the relative velocity of 1\%, then we arrive at an estimate $\epsilon \leq 10^{-40}\;{\rm kg\, s^2}$.  

\subsection{Interferometry}

Acceleration can be measured directly , e.g with atomic interferometry. This has been proposed first by Bord\'e \cite{Borde89} and the today's best performance gives an uncertainty of the measured acceleration of $\Delta \ddot x \approx 10^{-8} \; {\rm m/s^2}$ \cite{PetersChungChu99}. However, while this accuracy is valid for a constant acceleration, in our case we have a fast varying acceleration.

We use the phase shift 
\begin{equation}
\Delta \phi = A(\omega) \, k \, \ddot x \, T^2 \, ,
\end{equation}
with our acceleration \eqref{anomalousacceleration}. Since that part of our acceleration we are interested in and which we like to detect this way is fluctuating, we have to amend the standard phase shift for a dc acceleration, $\Delta\phi = k g T^2$, by a transfer function $A(\omega)$ which has been determined in \cite{Cheinetetal08}. We use as charged particle ionized Helium and take typical values for the laser wavelength $\lambda = 780\;{\rm nm}$, a pulse spacing time of $T = 100\;{\rm ms}$, an electric field strength of $E_0 = 10^{10}\;{\rm V/m}$. An experimentally reachable accuracy of the phase measurement is $\Delta\phi = 10^{-3} \; {\rm rad}$. 

We are interested in the largest $\omega$ which we are able to detect. With the specifications given we can determine that $\omega$ from the condition $A(\omega) = 10^{-25}$. This gives $\omega = 10^{12} \;{\rm Hz}$ which is the maximum frequency which effect on the phase shift we are able to detect. If we assume that nothing is detected this gives an estimate of 
\begin{equation}
\epsilon \leq \frac{m}{\omega^2} = 10^{-50} \, {\rm kg \, s^2} \, .
\end{equation}

\subsection{Electronic noise}

Another situation where a kind of {\it zitterbewegung} of a charged particle may induce an effect is electronics. As a most simple model we may assume that the particle under consideration is located within a capacitor. An oscillation of this particle will induce an electronic noise in the electric circuit. This electronic noise can be estimated by 
\begin{equation}
C \langle U^2 \rangle = m \langle \dot x^2 \rangle 
\end{equation}
where $U$ is the voltage. Using only the oscillating terms in the velocity we obtain 
\begin{equation}
C \langle U^2 \rangle = \frac{1}{2} \epsilon \left(\frac{q}{m} E_0\right)^2
\end{equation}
Therefore, a modified dynamics will induce an electronic noise -- beside the noise like Nyquist noise or a shot noise. However, due to its different characteristics, it might be disentangled from the other noise sources. Our noise should be a fundamental noise not depending on temperature, the finite number of charged particles in the electric circuit, etc. A good cryogenic noise limit is of the order 1 nV/$\sqrt{\rm Hz}$ \cite{HuYang05} in a wide frequency range, that is 1 nV for a measurement of duration 1 s. Taking as granted that no fundamental noise of this kind has been seen under the conditions of a molecular vacuum and cryogenic temperature, we may get an first estimate $\epsilon \leq 10^{-68}\;{\rm kg \, s^2}$, where we took a capacitance of $0.48 \; {\rm pF}$, a distance between the capacitor plates of $15 \; {\rm \mu m}$, a voltage of $1000 \; {\rm V}$. Taking into account a bandwith of 1 GHz, we get a more realistic estimate $\epsilon \leq 10^{-50} \;{\rm kg \, s^2}$.

%Aber auch die initial conditions kann man nut mit einer bestimmten Fehlermarge festlegen ... d.h. benoetige etwas, was vermoege von Dynamik Streuung in Streuung abbildet ... Lyapunov Exponent ?

%Liouvile theorem ---

%Flaeche im Phasenraum ---- Flaeche im Phasenraum

\section{Conclusion and outlook}

We studied the physics resulting from a hypothetically given higher order equation of motion substituting standard Newtons law. In order to confront such a model with experiments we first have to introduce the coupling to external force fields. This we employed via the standard gauge principle. We therefore started from a higher order Lagrangian and first introduced the interaction with external fields through a standard gauge principle. Since it is applied to higher order Lagrangians, the resulting gauge field have a richer structure than in the ordinary first order formalism. Beside the ordinary gauge fields usually obtained in first order formalism, we also obtained a space metric as a gauge field. As a by--product we obtained as a second order gauge field the standard space--time metric. 

Then we discussed physical consequences of equations of motion with higher order time derivatives coupled to these gauge field. We solved the equation of motion for the simplest case and deduced observational consequences in a few very simple experimental situations. Small higher order terms may influence the time of flight of accelerated particles, it yields a large acceleration fluctuation accessible in atomic interferometry, and also a fundamental noise induced in electronic devices. No deviation from, Newton's second law has been observed. Very rough and preliminary estimates for the parameter characterizing the higher order term could be derived. However, a comparison with the a Planck scale version of this parameter shows that the experimental estimates are far away from testing quantum gravity.  

The next step is to implement a higher order theory for point particles in a relativistic context. A possible Lagrange function to start with might be $L = m \sqrt{\eta_{ab} \dot x^a \dot x^b + \epsilon \eta_{ab} \ddot x^a \ddot x^b}$, where the dot is the derivative with respect to some parameter along the path. We also like to implement our principle in higher order field theory, e.g., for scalar field Lagrange densities of the type ${\cal L} = \eta^{ab} \partial_a\phi^* \partial_b \phi + \mu \eta^{ab} \eta^{cd} \partial_a\partial_c \phi^* \partial_b \partial_d \phi$. As further step would be to set up some field equations for the new gauge fields. A first guess might be to have the usual Maxwell equations for the $A_a$ and the Einstein field equations for the $A_{ab}$. 

\section{Acknowledgments}

We like to thank E. G\"okl\"u, S. Herrmann, J. Kunz, O. Lechtenfeld, and B. Mashhoon for discussions. C.\,L. also would like to thank the German Aerospace Center DLR for financial support. C.\,L. and P.\,R. also acknowledge financial support from the German Research Foundation and the Centre for Quantum Engineering and Space--Time Research QUEST.

%\bibliographystyle{unsrt}
%\bibliography{qtn}

\end{document}